%% file: SbBe_source_with_iron_filter.tex
\documentclass[%
 reprint,
preprintnumbers,
 amsmath,amssymb,
 aps,
prd,
 superscriptaddress,
]{revtex4-2}

\pdfoutput=1
\usepackage{siunitx}
\usepackage{graphicx}
\usepackage{dcolumn}
\usepackage{multirow}
\usepackage{bm}
\usepackage{hyperref}
\usepackage[mathlines]{lineno}
\linenumbers\relax 
\begin{document}

\preprint{APS/123-QED}
\nolinenumbers
\title{A portable and high intensity 24 keV neutron source based on $^{124}$Sb-$^9$Be photoneutrons and an iron filter}

\include{SPICE_HERALD_author-list}

\collaboration{SPICE/HeRALD Collaboration}

\date{\today}

\begin{abstract}
\input{Section_text/abstract.tex}
\end{abstract}

\maketitle

\section{Introduction}
\label{sec:Introduction}
\input{Section_text/introduction.tex}

\section{Source construction}
\label{sec:Source-construction}
\input{Section_text/construction.tex}

\section{Monte Carlo simulations}
\label{sec:simulations}
\input{Section_text/monte-carlo.tex}

\section{Source Characterization}
\label{sec:Source-Characterization}
\input{Section_text/characterization.tex}

\section{Neutron tagging with a liquid scintillator detector}
\label{sec:Neutron-tagging}
\input{Section_text/neutron_tagging.tex}

\section{Conclusion}
\label{sec:conclusion}
\input{Section_text/conclusion.tex}

\begin{acknowledgments}
This work was supported in part by DOE Grant DE-SC0019319, and DOE Quantum Information Science Enabled Discovery (QuantISED) for High Energy Physics (KA2401032). This material is based upon work supported by the National Science Foundation Graduate Research Fellowship under Grant No. DGE 1106400. This material is based upon work supported by the Department of
Energy National Nuclear Security Administration through the Nuclear Science and
Security Consortium under Award Number(s) DE-NA0003180 and/or DE-NA0000979. This research used the Savio computational cluster resource provided by the Berkeley Research Computing program at the University of California, Berkeley (supported by the UC Berkeley Chancellor, Vice Chancellor for Research, and Chief Information Officer). W. G. acknowledges the support from the National High Magnetic Field Laboratory at Florida State University, which is funded through the NSF Cooperative Agreement No. DMR-1644779 and the state of Florida. 
\end{acknowledgments}

\bibliography{Helium_Bib_12_2020.bib}

\end{document}

%% file: SPICE_HERALD_author-list.tex
\author{A.~Biekert} \affiliation{University of California Berkeley, Department of Physics, Berkeley, CA 94720, USA}
\author{C.~Chang} \affiliation{Argonne National Laboratory, 9700 S Cass Ave, Lemont, IL 60439, USA}
\affiliation{Department of Astronomy and Astrophysics, University of Chicago, Chicago, IL 60637, USA}
\affiliation{Kavli Institute for Cosmological Physics, University of Chicago, Chicago, IL 60637, USA}
\author{L.~Chaplinsky} \affiliation{University of Massachusetts, Amherst Center for Fundamental Interactions and Department of Physics, Amherst, MA 01003-9337 USA}
\author{C.~W.~Fink} \affiliation{University of California Berkeley, Department of Physics, Berkeley, CA 94720, USA} 
\author{W.~D.~Frey} \affiliation{McClellan Nuclear Research Center–UC Davis, McClellan, CA 95652, USA} 
\author{M.~Garcia-Sciveres} \affiliation{Lawrence Berkeley National Laboratory, 1 Cyclotron Rd., Berkeley, CA 94720, USA}
\author{W.~Guo} \affiliation{Department of Mechanical Engineering, FAMU-FSU College of Engineering, Florida State University, Tallahassee, FL 32310, USA} \affiliation{National High Magnetic Field Laboratory, Tallahassee, FL 32310, USA}
\author{S.A.~Hertel} \affiliation{University of Massachusetts, Amherst Center for Fundamental Interactions and Department of Physics, Amherst, MA 01003-9337 USA}
\author{X.~Li} \affiliation{Lawrence Berkeley National Laboratory, 1 Cyclotron Rd., Berkeley, CA 94720, USA}
\author{J.~Lin} \affiliation{University of California Berkeley, Department of Physics, Berkeley, CA 94720, USA}
\author{M.~Lisovenko} \affiliation{Argonne National Laboratory, 9700 S Cass Ave, Lemont, IL 60439, USA}
\author{R.~Mahapatra} \affiliation{Texas A\&M University, Department of Physics and Astronomy, College Station, TX 77843-4242, USA}
\author{D.~N.~McKinsey} \affiliation{University of California Berkeley, Department of Physics, Berkeley, CA 94720, USA} \affiliation{Lawrence Berkeley National Laboratory, 1 Cyclotron Rd., Berkeley, CA 94720, USA}
\author{S.~Mehrotra} \affiliation{University of California Berkeley, Department of Physics, Berkeley, CA 94720, USA}
\author{N.~Mirabolfathi} \affiliation{Argonne National Laboratory, 9700 S Cass Ave, Lemont, IL 60439, USA}
\author{P.~K.~Patel} \affiliation{University of Massachusetts, Amherst Center for Fundamental Interactions and Department of Physics, Amherst, MA 01003-9337 USA}
\author{B.~Penning} \affiliation{University of Michigan, Randall Laboratory of Physics, Ann Arbor, MI 48109-1040, USA} 
\author{H.~D.~Pinckney} \affiliation{University of Massachusetts, Amherst Center for Fundamental Interactions and Department of Physics, Amherst, MA 01003-9337 USA}
\author{M.~Reed} \affiliation{University of California Berkeley, Department of Physics, Berkeley, CA 94720, USA}
\author{R.~K.~Romani} \affiliation{University of California Berkeley, Department of Physics, Berkeley, CA 94720, USA}
\author{B.~Sadoulet} \affiliation{University of California Berkeley, Department of Physics, Berkeley, CA 94720, USA} 
\author{R.~J.~Smith} \affiliation{University of California Berkeley, Department of Physics, Berkeley, CA 94720, USA} 
\author{P.~Sorensen} \affiliation{Lawrence Berkeley National Laboratory, 1 Cyclotron Rd., Berkeley, CA 94720, USA} 
\author{B.~Suerfu} \affiliation{University of California Berkeley, Department of Physics, Berkeley, CA 94720, USA} 
 \author{A.~Suzuki} \affiliation{Lawrence Berkeley National Laboratory, 1 Cyclotron Rd., Berkeley, CA 94720, USA}
\author{V.~Velan} \affiliation{Lawrence Berkeley National Laboratory, 1 Cyclotron Rd., Berkeley, CA 94720, USA} 
\author{G.~Wang} \affiliation{Argonne National Laboratory, 9700 S Cass Ave, Lemont, IL 60439, USA}
\author{Y.~Wang} \affiliation{University of California Berkeley, Department of Physics, Berkeley, CA 94720, USA} 
\author{S.~L.~Watkins} \affiliation{University of California Berkeley, Department of Physics, Berkeley, CA 94720, USA}
\author{M.~R.~Williams} \affiliation{University of Michigan, Randall Laboratory of Physics, Ann Arbor, MI 48109-1040, USA} 

%% file: Section_text/abstract.tex
A portable monoenergetic 24~keV neutron source based on the $^{124}$Sb-$^9$Be photoneutron reaction and an iron filter has been constructed and characterized. The coincidence of the neutron energy from SbBe and the low interaction cross-section with iron (mean free path up to 29~cm) makes pure iron specially suited to shield against gamma rays from $^{124}$Sb decays while letting through the neutrons. To increase the $^{124}$Sb activity and thus the neutron flux, a $>$1~GBq $^{124}$Sb source was produced by irradiating a natural Sb metal pellet with a high flux of thermal neutrons in a nuclear reactor. The design of the source shielding structure makes for easy transportation and deployment. A hydrogen gas proportional counter is used to characterize the neutrons emitted by the source and a NaI detector is used for gamma background characterization. At the exit opening of the neutron beam, the characterization determined the neutron flux in the energy range 20-25~keV to be 5.36$\pm$0.20 neutrons per cm$^2$ per second and the total gamma flux to be 213$\pm$6 gammas per cm$^2$ per second (numbers scaled to 1~GBq activity of the $^{124}$Sb source). A liquid scintillator detector is demonstrated to be sensitive to neutrons with incident kinetic energies from 8 to 17~keV, so it can be paired with the source as a backing detector for neutron scattering calibration experiments. This photoneutron source provides a good tool for in-situ low energy nuclear recoil calibration for dark matter experiments and coherent elastic neutrino-nucleus scattering experiments. 

%% file: Section_text/introduction.tex
For dark matter direct detection experiments \cite{LUX:2019, LZ:2020NIM, Aprile_2020, SCDMS:2017, EDWIII:2016, CRESST-III:2019, PANDAX_4T:2021} and coherent elastic neutrino-nucleus scattering (CEvNS) experiments \cite{Ricochet_2017, COHERENT:2017, COHERENT:2021}, signal events consist of low energy nuclear recoils (less than tens of keV). Dark matter experiments like SPICE \cite{TESSERACT:2021} and HeRALD \cite{HERALD:2019} look for sub-GeV dark matter, with nuclear recoil (NR) signals at the sub-keV scale. Understanding the detector response to low energy nuclear recoils is thus critical for the exploration of new physics with these experiments.

One straightforward way to calibrate the detector NR response is to use neutron sources, which can come in various forms. Neutrons can be generated by nuclear reactions from accelerating charged particles in portable devices. A deuterium-deuterium (DD) neutron generator produces monoenergetic neutrons with kinetic energies of about 2.5~MeV \cite{LUX_DD:2016}. Combined with a deuterium reflector \cite{D_reflector:2017}, the DD generator can produce 272~keV quasi-monoenergetic neutrons, but the neutron flux is decreased by a factor of 100. A deuterium-tritium (DT) neutron generator produces monoenergetic neutrons at 14~MeV \cite{knoll2010radiation}. Neutrons can also be generated through ($\alpha$,n) nuclear reactions, for example with AmLi \cite{TAGZIRIA20122395} or AmBe \cite{VIJAYA1973435} sources. Neutrons from spontaneous fission of $^{252}$Cf \cite{Mannhart1989} are also commonly used. However, neutrons from ($\alpha$,n) reactions or spontaneous fission are produced with continuous energy spectra. The most probable neutron energies are 140~keV for AmLi, $\approx$2~MeV for AmBe, and $\approx$2~MeV for $^{252}$Cf. Calibration with monoenergetic neutron sources is generally easier to interpret and produces lower systematic uncertainty than with continuous neutron energy spectra. With monoenergetic neutrons, one can determine the nuclear recoil energy deposit in the target nucleus on an event-by-event level by tagging the outgoing angle of the neutron; see Ref.\cite{HeRALD:2022} as an example. 

Ultra-low energy ($<$ 1~keV) nuclear recoil calibration can also be achieved with a gamma source by coherent photon-nucleus scattering \cite{Robinson:2017}. Additionally, calibration with sub-keV nuclear recoils induced by neutron capture has been demonstrated at 112~eV in tungsten nuclei \cite{Abele:2022}. 

The photonuclear reaction $^9$Be($\gamma$, n)$^8$Be is another option to produce quasi-monoenergic neutrons. $^{88}$Y-$^9$Be (154~keV neutron) \cite{Collar:2013} and $^{124}$Sb-$^9$Be (24~keV neutron) \cite{Chavarria:2016, Amole:2019} sources have been deployed to calibrate low energy rare event searches.  In these experiments, lead shielding was used to attenuate the gamma flux, which is generally several orders of magnitude higher than the neutron flux given the relatively small cross section for the photonuclear reaction. These examples of calibration with photoneutron sources rely on end-point calibration because the neutron flux is usually too low for neutron tagging with a backing detector. This is because the backing neutron detector only tags neutrons scattered into a small solid angle.

Larger facilities could also provide monoenergetic neutrons at low energy. The 10~MV tandem accelerator at Triangle Universities Nuclear Laboratory can generate monoenergetic neutrons via nuclear reactions with accelerated charged particles \cite{TUNL:2017}. For example, neutrons from the $^7$Li(p,n)$^7$Be reaction can be tuned to energies between 70~keV and 650~keV. Besides particle accelerators, nuclear reactors also generate neutrons. A filtered neutron source using an iron filter at a nuclear reactor has been demonstrated for 24~keV monoenergetic neutrons \cite{BARBEAU2007385}. Experiments need to relocate to these facilities to utilize these low energy neutrons, making these calibrations more useful to study the intrinsic properties of the detector medium (e.g. intrinsic charge yield and light yield of liquid noble elements \cite{Lenardo:2019}). But this ex-situ calibration does not capture detector-specific response, as the detectors are usually operated at other research facilities. The detector-specific response is critical for dark matter and CEvNS searches.

An ideal neutron calibration source for dark matter and CEvNS experiments should be monoenergetic and portable, with a high flux of low energy neutrons and low backgrounds. A portable, robust and high-flux neutron source is also useful for neutron activation studies for security applications \cite{GREENBERG2011193}. 

With these goals in mind, we present a portable 24~keV neutron source based on $^{124}$Sb-$^9$Be photoneutrons and an iron filter in this article.

%% file: Section_text/construction.tex
\begin{figure}[t]
\includegraphics[width=\linewidth]{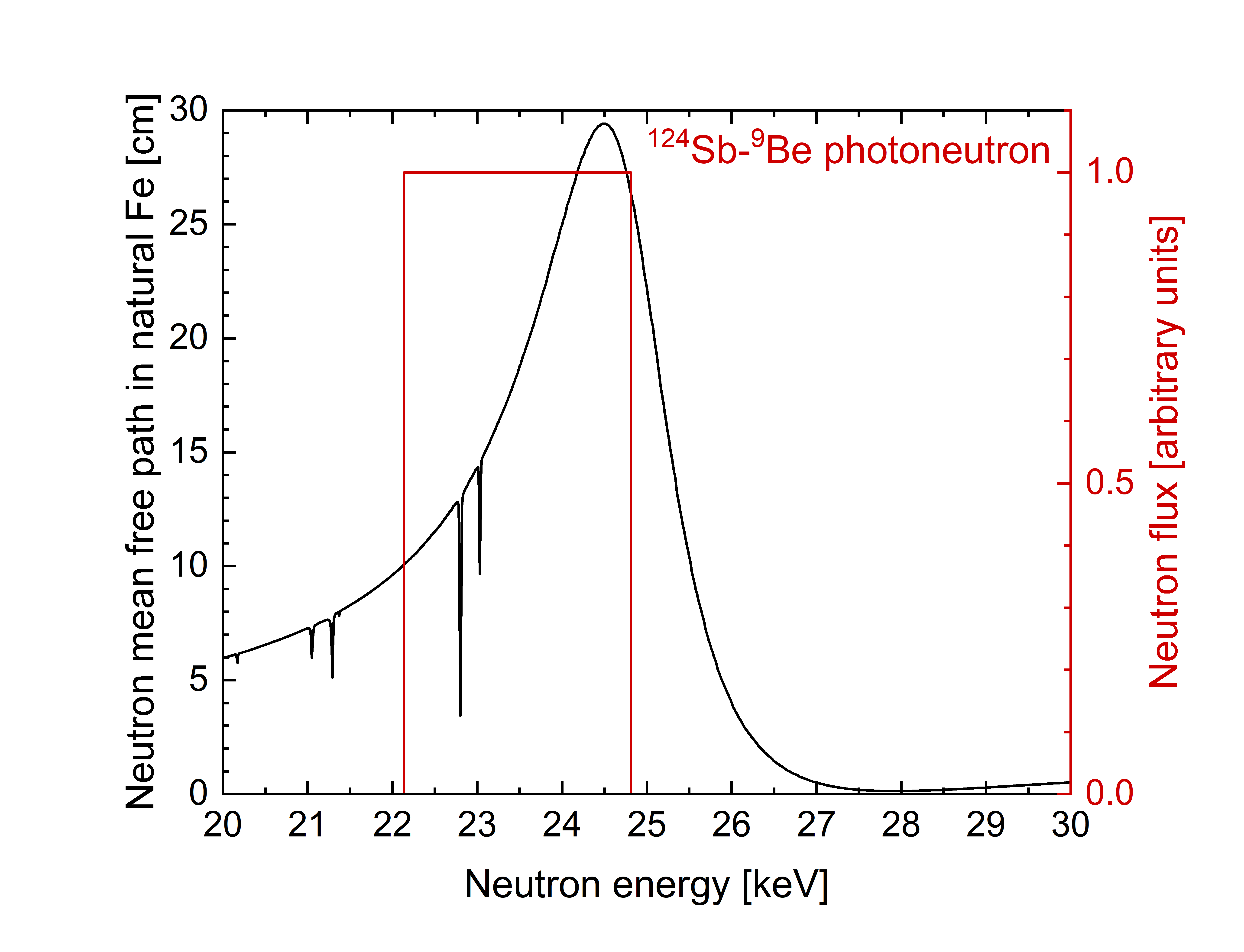}
\caption{\label{fig:SbBe_Fe_coincidence} 
The neutron spectrum of $^{124}$Sb-$^9$Be photoneutrons and the neutron mean free path in natural iron. The neutrons generated by the SbBe source coincide with the crest in the neutron mean free path in iron. Neutron interaction cross section (thus mean free path) data is from ENDF/B-VIII.0 database \cite{DB2018a}.}
\end{figure}

As shown in Fig. ~\ref{fig:SbBe_Fe_coincidence}, the neutron spectrum produced in the $^{124}$Sb-$^9$Be reaction coincides with a maximum of neutron mean free path in natural iron. In other words, iron is very transparent (mean free path up to 29~cm) to 24~keV SbBe neutrons while it remains an effective gamma shield. The energy of the neutron produced in the photoneutron reaction depends on the angle between the outgoing neutron and the incoming gamma in the reaction \cite{knoll2010radiation}, causing the spread in energy. One issue for photoneutron sources in general is the low ratio of neutrons to gammas, with $\mathcal{O}(1)$ neutron produced per 10$^4$ gammas emitted in decays of the gamma source. The iron filter can help to improve this neutron-to-gamma ratio in the flux escaping the source assembly.

The $^{124}$Sb source (half-life 60.2~days) was produced by the irradiation of a high purity 
($>$99.999\%) bare metallic antimony pellet (6.4~mm diameter and 6.4~mm height cylinder) inside of the University of California Davis / McClellan Nuclear Research Center TRIGA reactor.  The central facility position, which has the highest flux in the reactor core ($\approx$10$^{13}$ thermal neutrons/cm$^2$/sec), was utilized to provide a source of sufficient radioactivity and minimal size. The antimony pellet was hermetically sealed in two nested titanium cylinders purged with nitrogen gas in order to minimize the risk of unwanted oxidation of the pellet. Direct assaying of the final activity of the pellet was not feasible due to the high source activity. Instead a smaller Sb calibration pellet was irradiated at the same reactor power level for a shorter duration, at the same temperature to account for thermal and resonance absorption as well as any temperature effects on cross sections. This weaker calibration pellet was assayed using gamma ray spectroscopy with a high purity germanium detector. A simple ratio of the irradiation times and weight of the two pellets was used to deduce the final activity of the main $^{124}$Sb source pellet at the end of irradiation.  The source pellet was irradiated for 81.2~hours at 1~MW reactor power over approximately two weeks to produce an end of irradiation $^{124}$Sb activity of 120~mCi (4.4~GBq). $^{122}$Sb (half-life 2.7~days) is also produced during the neutron activation, so the activated Sb pellet was suspended inside the reactor water pool but away from neutrons for longer than a month for the decay of the $^{122}$Sb. The source strength was at 1.5~GBq when extracted from the reactor water pool. Due to the relatively short half-life of $^{124}$Sb, the activity of the source changed noticeably during the characterization campaign, and characterization data presented in this article is normalized to 1~GBq $^{124}$Sb activity. The decay gamma information for $^{124}$Sb and the corresponding SbBe photoneutrons can be found in Table \ref{table:1}.

\begin{table}[t]
\centering
\begin{tabular}{| p{1.25cm} |p{1.3cm} | p{1.3cm}| p{2cm}| p{2cm}|}
 \hline
 Gamma energy & Intensity & Neutron energy & Neutron production cross section  & Normalized neutron production rate\\
 (keV) &  & (keV) & (mb) & \\ [0.5ex] 
 \hline
 602.7 & 97.8\% & N/A & N/A & N/A\\ 
 645.9 & 7.42\% & N/A & N/A & N/A\\
 722.8 & 10.8\% & N/A & N/A & N/A \\
 1691.0 & 47.79\% & 24 & 1.41 & 1\\
 2090.9 & 5.51\% & 379 & 0.412 & $3.36 \times 10^{-2}$ \\ [1ex] 
 \hline
 \end{tabular}
 \caption{The gamma radiations from $^{124}$Sb decay and the neutron production from the SbBe photonuclear reaction. Gamma radiations with less than 3\% intensity are not listed. The normalized neutron production rate is the product of gamma intensity and the corresponding neutron production cross section, normalized to the value of the 24~keV neutron. The gamma energy threshold for photoneutron production in $^9$Be is 1666~keV \cite{knoll2010radiation}. Cross sections are from ENDF/B-VIII.0 \cite{DB2018a}}
 \label{table:1}
\end{table}

Along with the $^{124}$Sb source, photoneutron production requires beryllium. In this work, Be metal disks with thin (\SI{20}{\micro\metre}) Ni plating are used, which increases the neutron production rate by $\approx$70\% compared to using beryllium oxide with the same physical dimensions. The thin Ni plating encapsulating the Be metal disk prevents the release of Be surface metal or oxide particles into the air, which could be detrimental to health. The Be metal parts consisted of two 2.54~cm diameter, 1.27~cm thick disks, sandwiching the Sb pellet from the top and bottom. A middle Be disk with a central hole for the Sb pellet is also added; this middle disk has 2.54~cm outer diameter, 7.4~mm inner diameter and is 6.6~mm thick.

\begin{figure}[ht]
\includegraphics[width=\linewidth]{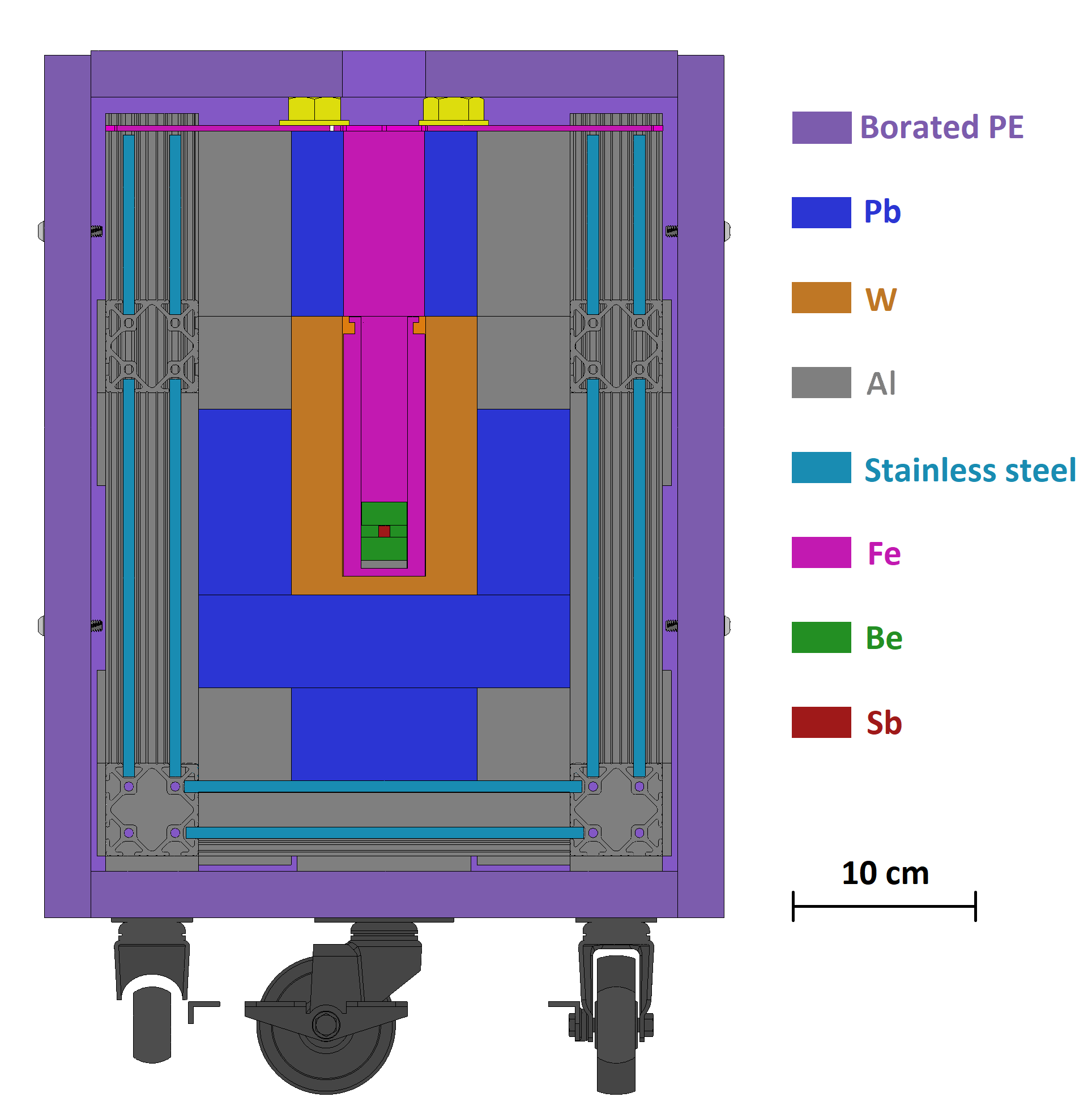}
\caption{\label{fig:SbBe_CAD_model} The CAD model cross section for the SbBe photoneutron source with color code for the materials. The red square in the center shows the activated Sb pellet}
\end{figure}

\begin{figure}[ht]
\includegraphics[width=\linewidth]{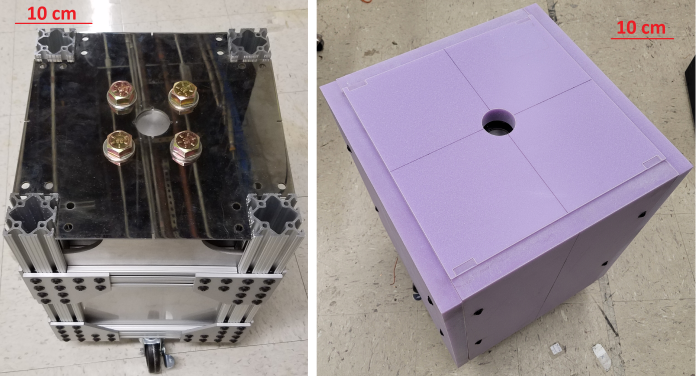}
\caption{\label{fig:SbBe_photo} Photos of the SbBe photoneutron source assembly. Left, source assembly before installing the borated PE panels. This is the transport configuration inside a 30-gallon drum. Right, source assembly with the borated PE panels installed. The central hole is for the neutron beam}
\end{figure}

A CAD model and photo of the source are shown in Fig.~\ref{fig:SbBe_CAD_model} and Fig.~\ref{fig:SbBe_photo}. The diameter of the central high purity ($>$99.9\%) iron rod is 4.47~cm. The total iron shielding length from the top surface of the beryllium metal disk to the top exit surface is 20.3~cm. 

A combination of tungsten, lead, stainless steel and aluminum materials are enclosed inside aluminium \mbox{T-slotted} framing rail as shielding structure for the source. Four casters are attached to the bottom of the framing rail for portability. This shielding structure weighs about 250 kg and can be transported inside an industrial 30 Gallon steel drum. The total weight of the source is designed to be below the drum tested weight limit as a Type-A radioactive material transport container, for the ease of transportation. After the source is transported to our laboratory from the nuclear reactor center, a layer of 2.54~cm thick 5\% borated polyethylene (PE) with a central opening for neutron beam was added to surround the source. This layer of borated PE decreases the stray neutron flux  from the sides and bottom, reducing the neutron radiation dose to personnel and suppressing the undesirable excess flux of neutrons with a broad, degraded energy spectrum scattering in the laboratory.

%% file: Section_text/monte-carlo.tex
The source geometry was simulated with \textsc{Geant4} \cite{Agostinelli:03nimpra, Allison:06itns, Allison:16nimpra} version~10.5 using the \textsc{Shielding} physics list. While most of the volumes were implemented exactly corresponding to the CAD model in Fig.~\ref{fig:SbBe_CAD_model}, the outer aluminum T-slotted framing rail was instead modeled as a solid volume with density equivalent to the volume-averaged aluminum density in the T-slot geometry, and the wheels were omitted completely.


The 24~keV neutrons of interest are produced from the $^{124}$Sb 1690.98~keV gamma ray line, which has an intensity of 47.79\% \cite{HI1997a}. For this gamma ray energy, the ENDF/B-VIII.0 photonuclear cross section \cite{DB2018a} is in good agreement with alternative fits from Robinson~\cite{AR2016c} and Arnold \textit{et al.}~\cite{CA2012a} with a value of 1.41~mb. We estimate the 24~keV neutron production rate to be $10^{-4}$ neutrons per $^{124}$Sb decay for the beryllium metal geometry used in this source.

\begin{figure}[ht]
\includegraphics[width=\linewidth]{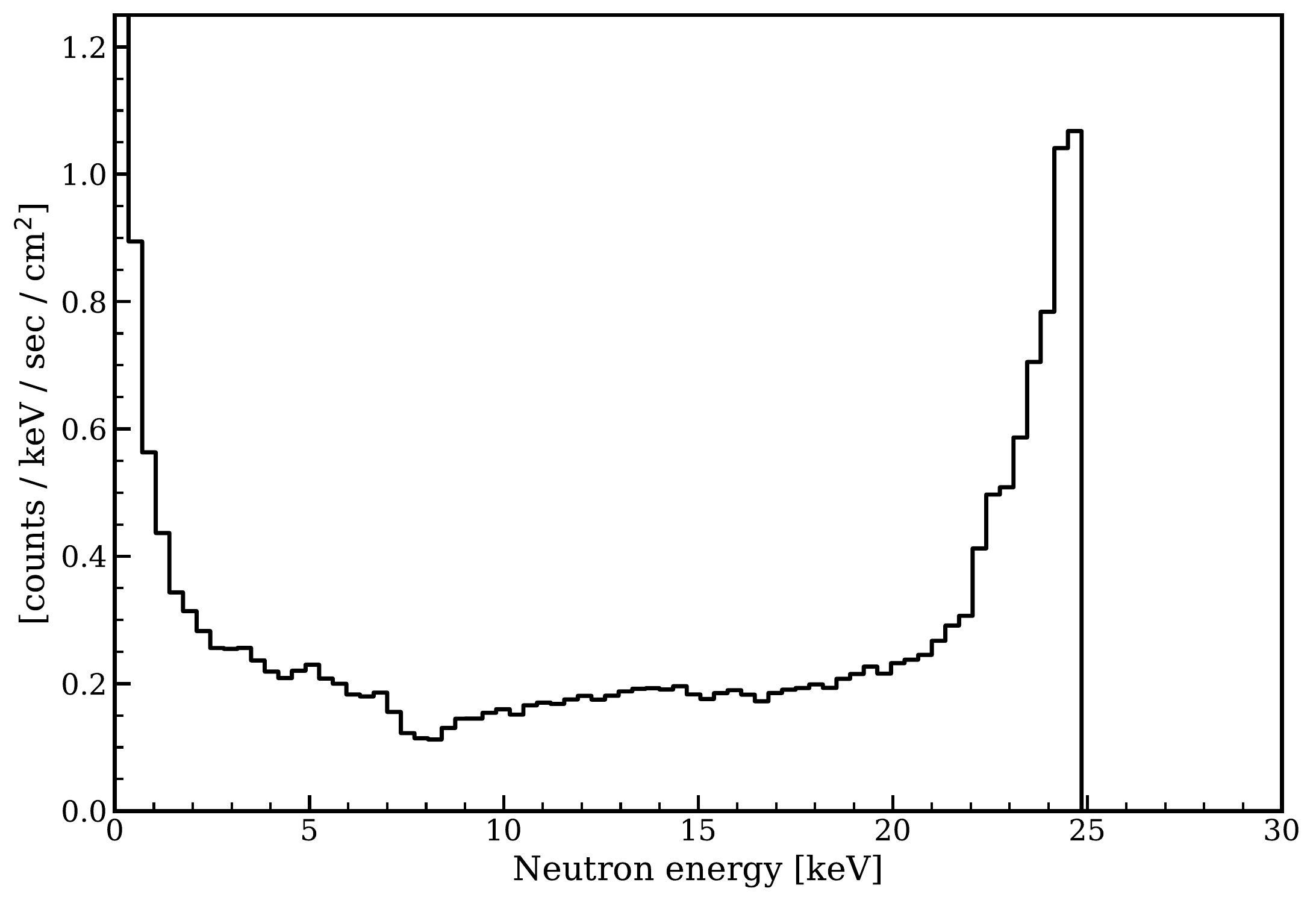}
\caption{\label{fig:sim_neutron_spectrum} The simulated neutron flux leaving the front source assembly opening, assuming a 1~GBq $^{124}$Sb source. Shown here is the spectrum associated with only the 24~keV neutrons most commonly produced in $^{124}$Sb decays.}
\end{figure}

The low energy neutron spectrum exiting the opening in the borated PE volume at the end of the iron filter is shown in Fig.~\ref{fig:sim_neutron_spectrum}. The flux of neutrons with energy $>$1~keV is 6.2~neutrons~per~cm$^2$~per~second assuming a 1~GBq $^{124}$Sb source. 40\% of this flux lies in the peak between 20 and 25~keV, indicating a fairly pure beam of low energy neutrons leaving the filter. The filter face has an area of 15.7~cm$^2$, yielding a 20-25~keV neutron emission rate of 38.4 neutrons per second. 

$^{124}$Sb has one additional significant decay gamma that should be considered in the context of photoneutron production, with an energy of 2090.95~keV and an intensity of 5.51\% \cite{HI1997a}. The neutron energy corresponding to these gamma rays is 379~keV. For the purposes of a monoenergetic, low energy neutron calibration, this neutron population represents a background. However, its production is subdominant not only because of the smaller branching fraction in $^{124}$Sb. At this gamma ray energy, the photonuclear cross section is reduced to 0.41~mb according to ENDF/B-VIII.0, and the values from Robinson~\cite{AR2016c} and Arnold \textit{et al.}~\cite{CA2012a} are even smaller at 0.26 and 0.25~mb, respectively. Considering the ENDF/B-VIII.0 value as a conservative estimate on the contamination from higher energy neutrons, we estimate that the 379~keV population accounts for only about 3\% of the total photoneutron production in the source and 5\% of the neutron flux leaving the front face of the iron filter. There are some other transitions above the photonuclear threshold in $^{124}$Sb decays, but their sub-percent level intensities are small enough that we do not consider them further.

\begin{figure}[ht]
\includegraphics[width=\linewidth]{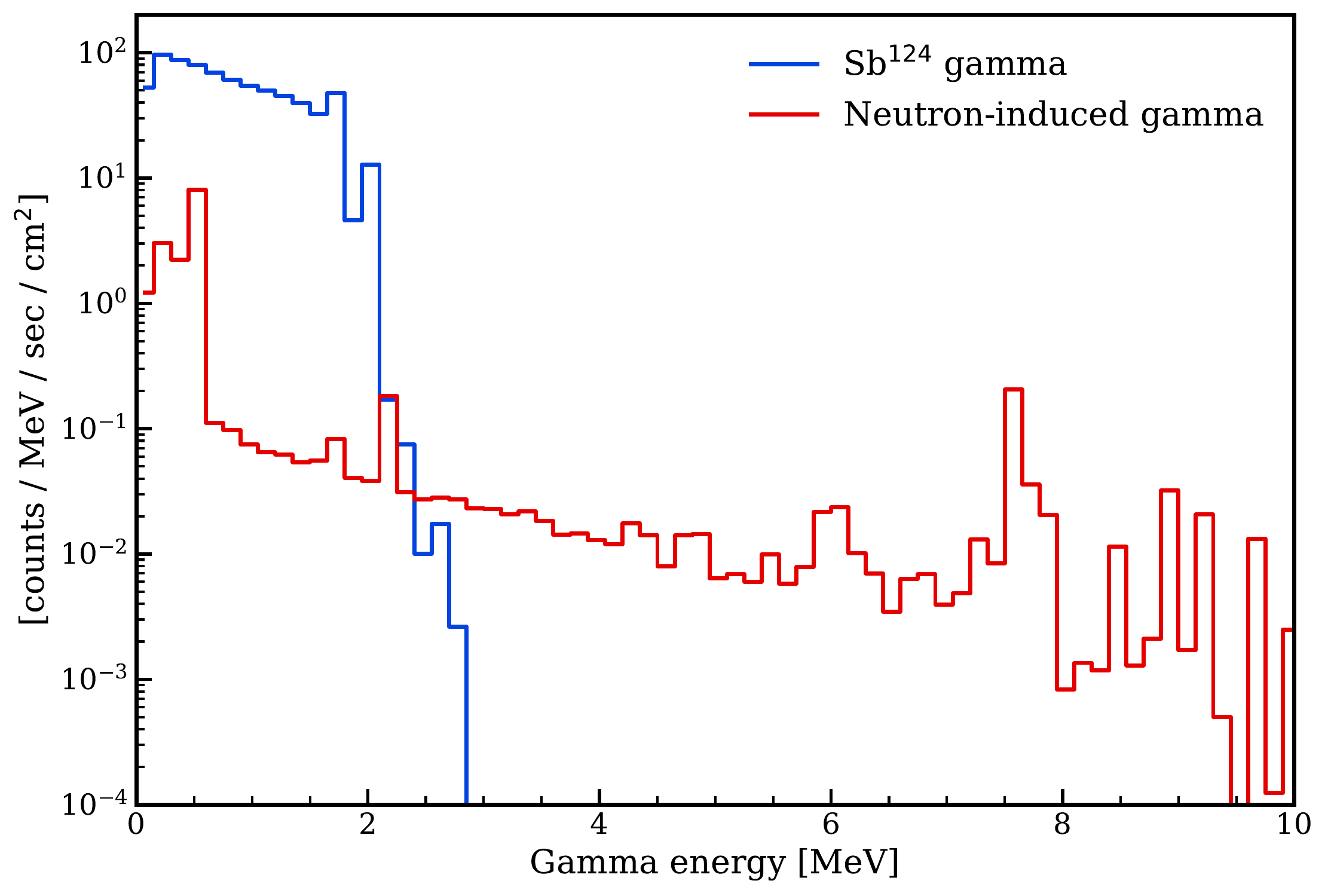}
\caption{\label{fig:sim_gamma_spectrum} The simulated gamma flux leaving the front face of the source assembly, assuming a 1~GBq $^{124}$Sb source. The spectrum is dominated by gamma rays originating from $^{124}$Sb decays, in blue. The spectrum in red is those gammas which originated from neutron capture or inelastic scattering off source materials.}
\end{figure}

As previously discussed, dense shielding material such as tungsten and lead reduces the intensity of gamma rays propagating into the experimental environment. In principle, neutrons can inelastically scatter and capture on these materials to produce extra gamma rays. Additionally, gamma rays from neutron capture can be produced in the borated PE that is added to capture off-axis neutrons. Fig. \ref{fig:sim_gamma_spectrum} shows that these secondary gammas originating from photoneutrons are subdominant to those gammas produced directly from $^{124}$Sb decays. The total flux leaving the front face of the source amounts to 108~gamma~rays~per~cm$^2$~per~second due to the dense shielding material in the line of sight between the $^{124}$Sb source and most of the front face. Since iron is not as dense as lead or tungsten, it is not as effective at shielding gamma rays, leading to a gamma ray flux at the end of the iron filter about 2.5 times larger than the flux averaged over the whole front face. The main consideration for a neutron calibration application is attenuating the total gamma ray flux enough to observe the low energy neutron spectrum. The acceptable ratio of neutron flux to gamma flux is dependent on the specific setup of the experiment as well as the ability to mitigate gamma backgrounds through techniques like time-of-flight, pulse shape discrimination, or background subtraction. The design simulated here is expected to be sufficient for a neutron scattering experiment, but the amount of shielding could be increased to further reduce the gamma ray flux if necessary for other applications.

%% file: Section_text/characterization.tex
Characterization of the SbBe source was performed using a hydrogen gas proportional counter (HGPC) for measuring neutrons and a NaI detector for measuring gammas.

\begin{figure}[t]
\includegraphics[width=\linewidth]{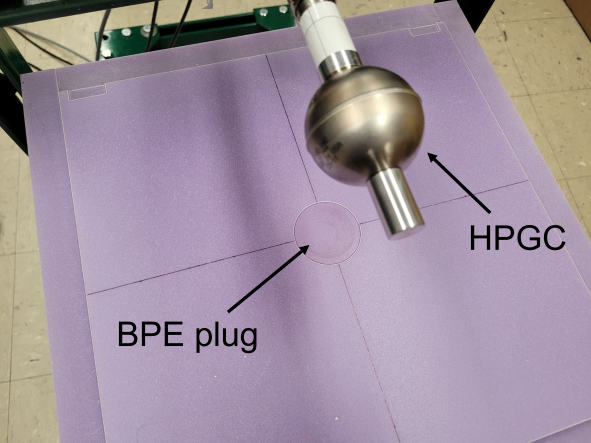}
\caption{\label{fig:proton_recoil_setup} Photo of the HGPC as placed for on-axis measurement of the SbBe photoneutron spectrum. In this image, the removable borated PE plug is in place to measure the gamma background.}
\end{figure}

A spherical hydrogen gas proportional counter, LND Model 27044 with 4.98~cm effective diameter and 3.04~bar hydrogen gas at 21 °C was deployed for measurement of the neutron energy spectrum. The close matching of the masses between the neutron and the proton allows for neutrons to deposit up to nearly 100\% of their energy in elastic scattering off of nuclei in this detector, which is valuable for measuring small neutron energies. For measurement of the on-axis neutron spectrum, the HGPC was placed 5.45~cm above the top borated PE surface of the source assembly, as shown in Fig.~\ref{fig:proton_recoil_setup}. A removable borated PE plug is also indicated in this photo, which was used for gamma background subtraction as described below.

Given the particular dimensions and pressure of the HGPC, incident gammas often deposit energies in this detector that are similar to those from the neutrons of interest, so that background must be considered when determining the neutron spectrum. Two strategies were taken to handle the background due to gammas. Firstly, gamma interactions produce recoiling electrons which have tracks in the hydrogen gas that, for energies of tens of keV or more, are centimeters long. The recoiling protons from elastic neutron scattering on the other hand are less than 1~mm for energies below 50~keV. This allows for pulse shape discrimination, since the recoiling electrons may produce ionizations at widely varying radius in the detector, causing the rise time distribution for pulses due to electron recoils to extend higher than for low energy nuclear recoils. Secondly, the removable borated PE plug placed at the end of the iron filter can be added to minimize the neutron flux and separately measure the gamma background. 

\begin{figure}[ht]
\includegraphics[width=\linewidth]{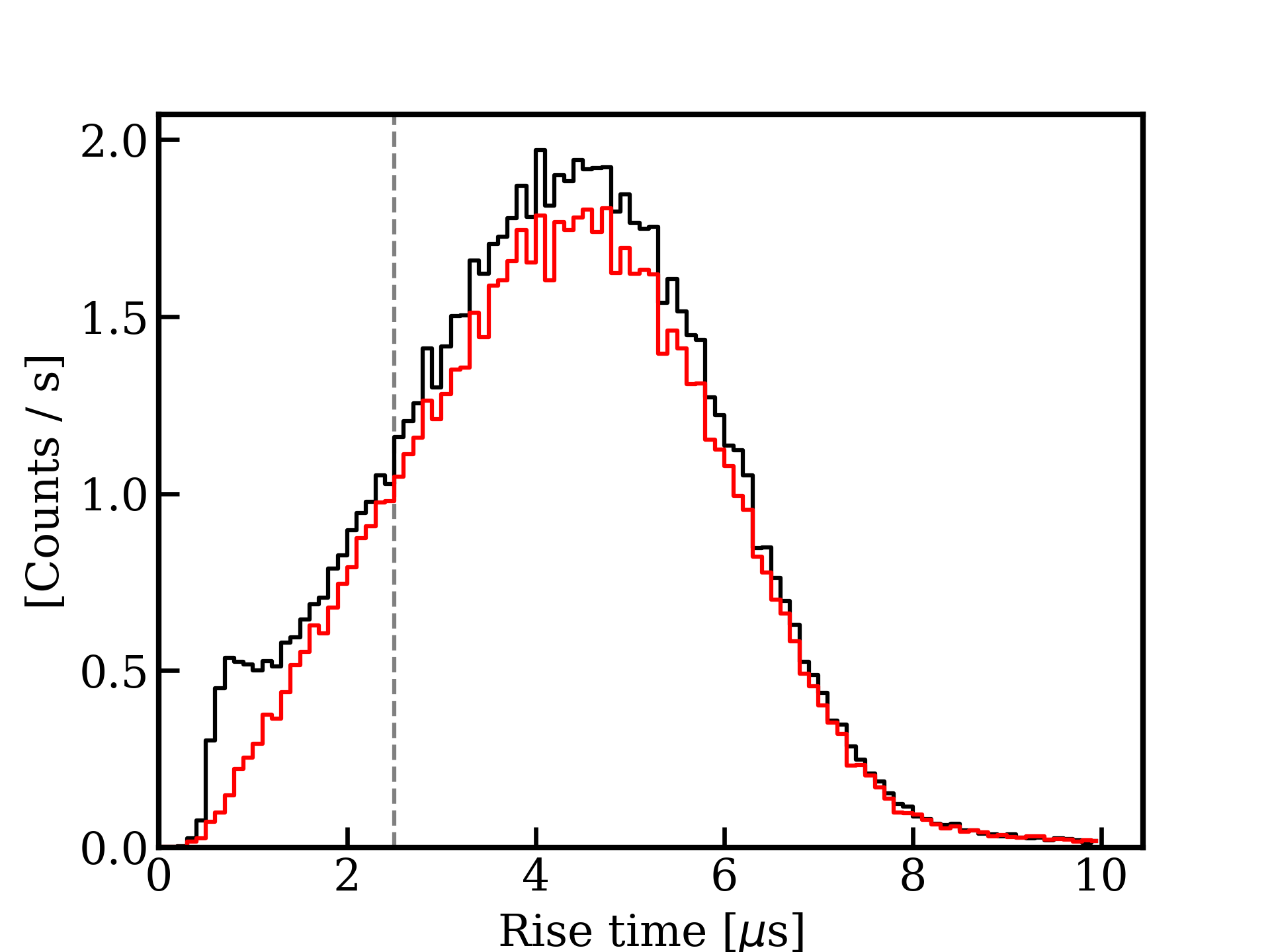}
\caption{\label{fig:proton_recoil_risetime} The rise time, defined as time from 10\% to 90\% pulse height, of pulses in the HGPC with (red) and without (black) a borated PE plug over the iron filter. The dashed line indicates the cut value of 2.5~$\mu$s. Only pulses with rise time shorter than this were retained, in order to reduce gamma background.}
\end{figure}

\begin{figure}[ht]
\includegraphics[width=\linewidth]{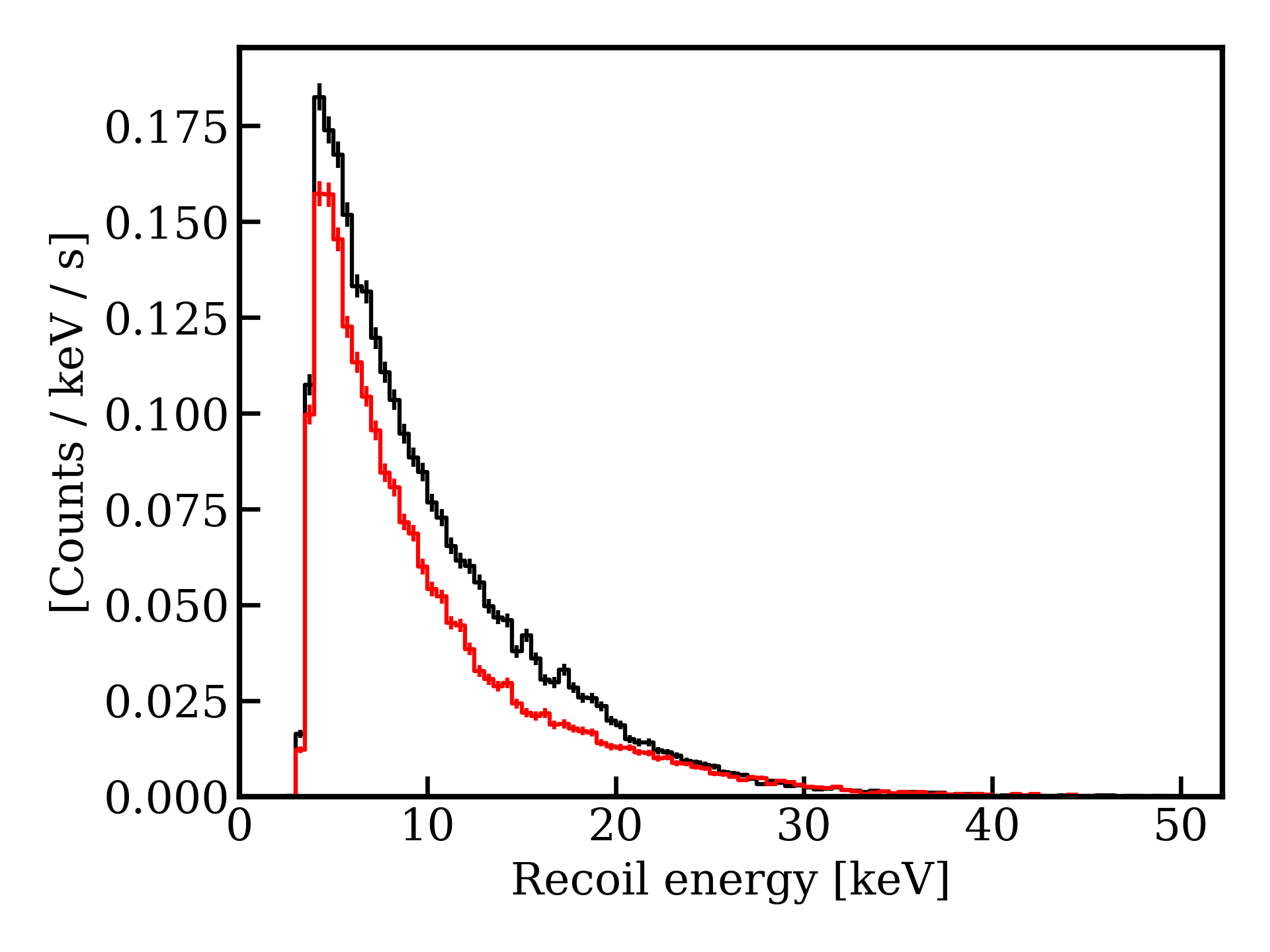}
\caption{\label{fig:proton_recoil_spectra} The recoil energy spectrum in the HGPC placed on-axis over the iron filter, with (red) and without (black) a borated PE plug over the iron filter, after the cut on rise time of the HGPC pulses.}
\end{figure}

The rise time distribution of pulses in the HGPC with and without the borated PE plug is shown in Fig.~\ref{fig:proton_recoil_risetime}. A clear excess at small rise times is present without the borated PE plug in place. To increase neutron purity in the pulse height spectrum, pulses with rise time greater than 2.5~$\mu$s are removed. Above this cut value the ratio of the rise time spectra with and without the borated PE plug is flat, and adjusting the cut value upwards does not increase the total rate in the final subtracted spectrum, indicating that this choice preserves the full neutron flux. The resulting recoil energy spectra are shown in Fig.~\ref{fig:proton_recoil_spectra}; the horizontal axis is placed in units of energy by a constant of proportionality between deposited energy and pulse height in the HGPC. 

\begin{figure}[ht]
\includegraphics[width=\linewidth]{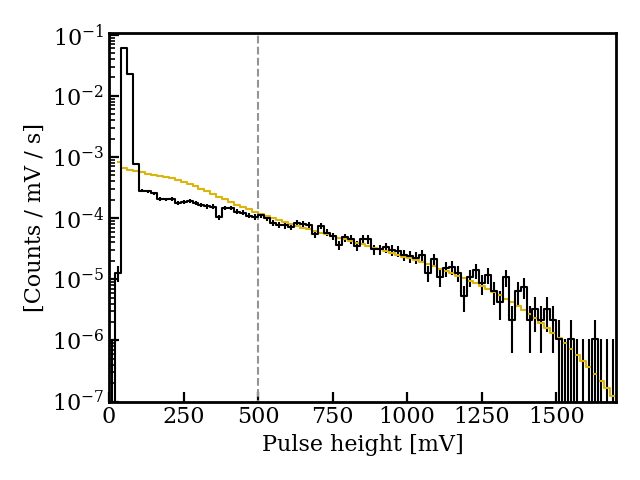}
\caption{\label{fig:proton_recoil_DD} The pulse height spectrum in the HGPC when exposed to a deuterium-deuterium neutron generator, along the acceleration axis of the generator in the forward direction. This is used to calibrate the energy scale and resolution of the HGPC by fitting the data (black) to simulation (gold). The data left of the dashed line is not included in the fit.}
\end{figure}

The constant of proportionality is determined from the best fit of the \textsc{Geant4} simulation to data taken with a DD neutron generator. A linear resolution parameter is also allowed to vary in the fit, and thereby measured to be $10\pm3$\%. The best fit to the DD calibration data is shown in Fig.~\ref{fig:proton_recoil_DD}. Note that the 2.8~MeV DD neutrons are much higher energy than the SbBe photoneutrons, and protons recoiling with MeV-scale energy in the HGPC have track length comparable to the detector dimensions. Therefore the shape of the spectrum is strongly influenced by the detector geometry, and does not exhibit a sharp endpoint. To better fit the smooth roll-off of the recoil spectrum from DD neutrons, a lower bound to the fit window is placed at pulse heights of 500~mV (which in the best fit corresponds to 275~keV), below which the observed pulse height spectrum cannot be fit well by the simulation. Lower energies in the DD spectrum correspond to tracks that traverse smaller chords in the detector, so the failure to simulate this portion of the spectrum may indicate nonuniformity in the HGPC response near its edges.

\begin{figure}[ht]
\includegraphics[width=\linewidth]{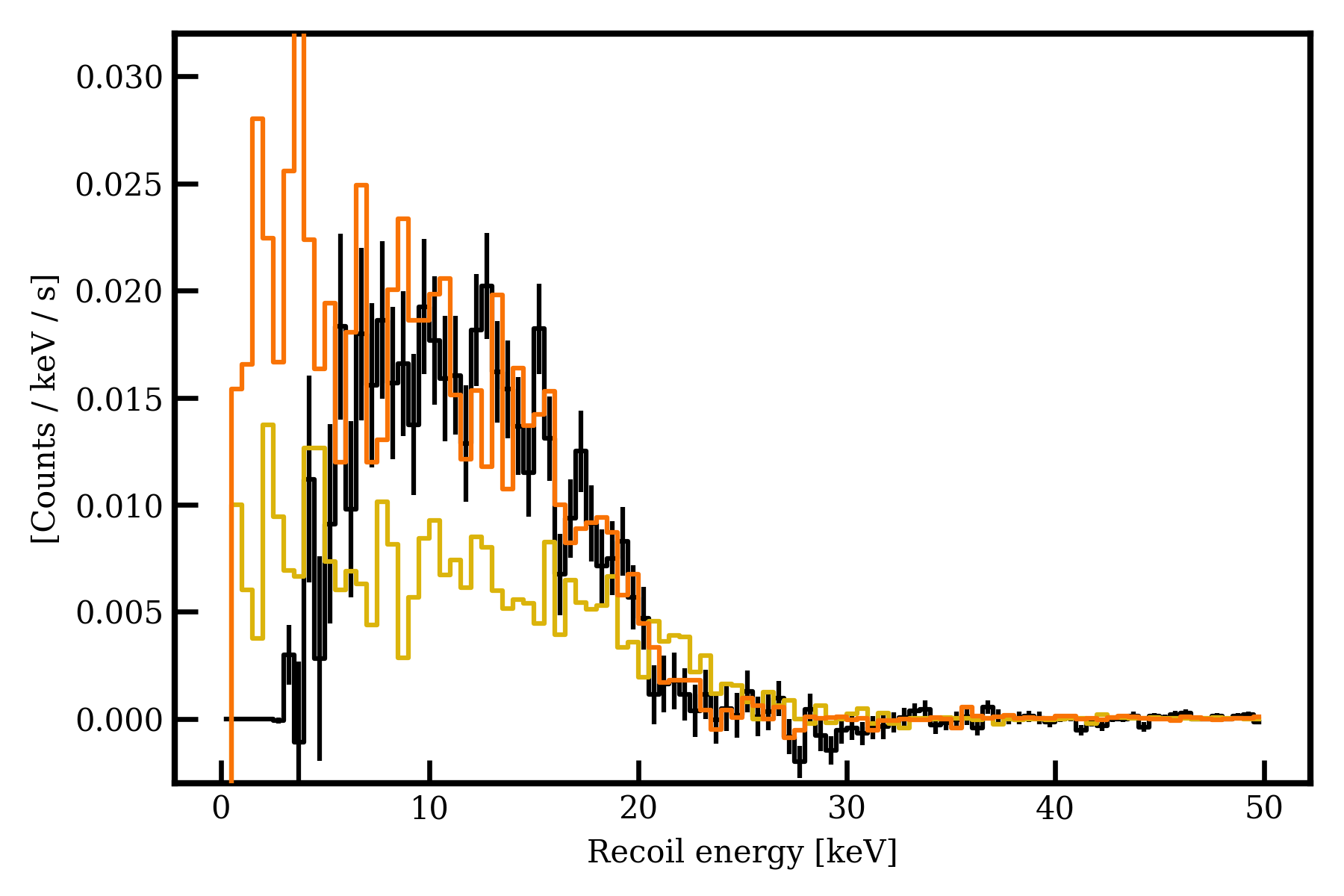}
\caption{\label{fig:proton_recoil_subtracted} The gamma-subtracted recoil energy spectrum in the HGPC (black). The expected spectrum from Monte Carlo simulation is shown in gold, while a best fit scaling of the simulation is in orange. The horizontal axis is scaled according to the DD calibration shown in Fig.~\ref{fig:proton_recoil_DD}.}
\end{figure}

To obtain the final neutron-only recoil energy spectrum, the gamma background measured with the borated PE plug in place is subtracted from the full spectrum without the borated PE plug. A gamma attenuation factor from the borated PE plug is calculated from the ratio of long rise time ($>$2.5~$\mu$s) pulses with and without the plug. The spectrum with borated PE plug in place is corrected by this factor prior to background subtraction. The resulting neutron-only recoil spectrum is shown in Fig.~\ref{fig:proton_recoil_subtracted}, in comparison with the spectrum from \textsc{Geant4} simulation, with no free parameters. In addition, a best fit spectrum is included in which the flux and energy scaling are allowed to float to fit the data down to 6~keV (below which the HGPC threshold complicates comparison with simulation). The observed spectrum has a fairly distinct endpoint, but it lies closer to 20~keV than the expected 24~keV; the best fit indicates that the simulation better agrees with data if the simulated recoil energies are scaled by 0.84. It is likely this discrepancy could be attributed to a systematic error in the gain calibration using DD neutrons, perhaps due to an intrinsic deviation from proportionality in the HGPC between the disparate DD and photoneutron energy scales. The best fit scaling of the simulation also indicates that the observed rate is 2.16$\pm$0.12 above the simulated rate. With flux and energy scaling floated, the simulation fits the data with $\chi^2/d.o.f. = 32.6/48$. This indicates that the shape of the recoil spectrum is consistent between data and the fitted simulation, down to the 6~keV threshold and that therefore the data is consistent with the simulated neutron spectrum exiting the source. Assuming that the simulation accurately captures the neutron spectrum shape, the flux of 20-25~keV neutrons exiting the hole in the borated PE is 5.36$\pm$0.20 neutrons per cm$^2$ per second.

\begin{figure}[ht]
\includegraphics[width=\linewidth]{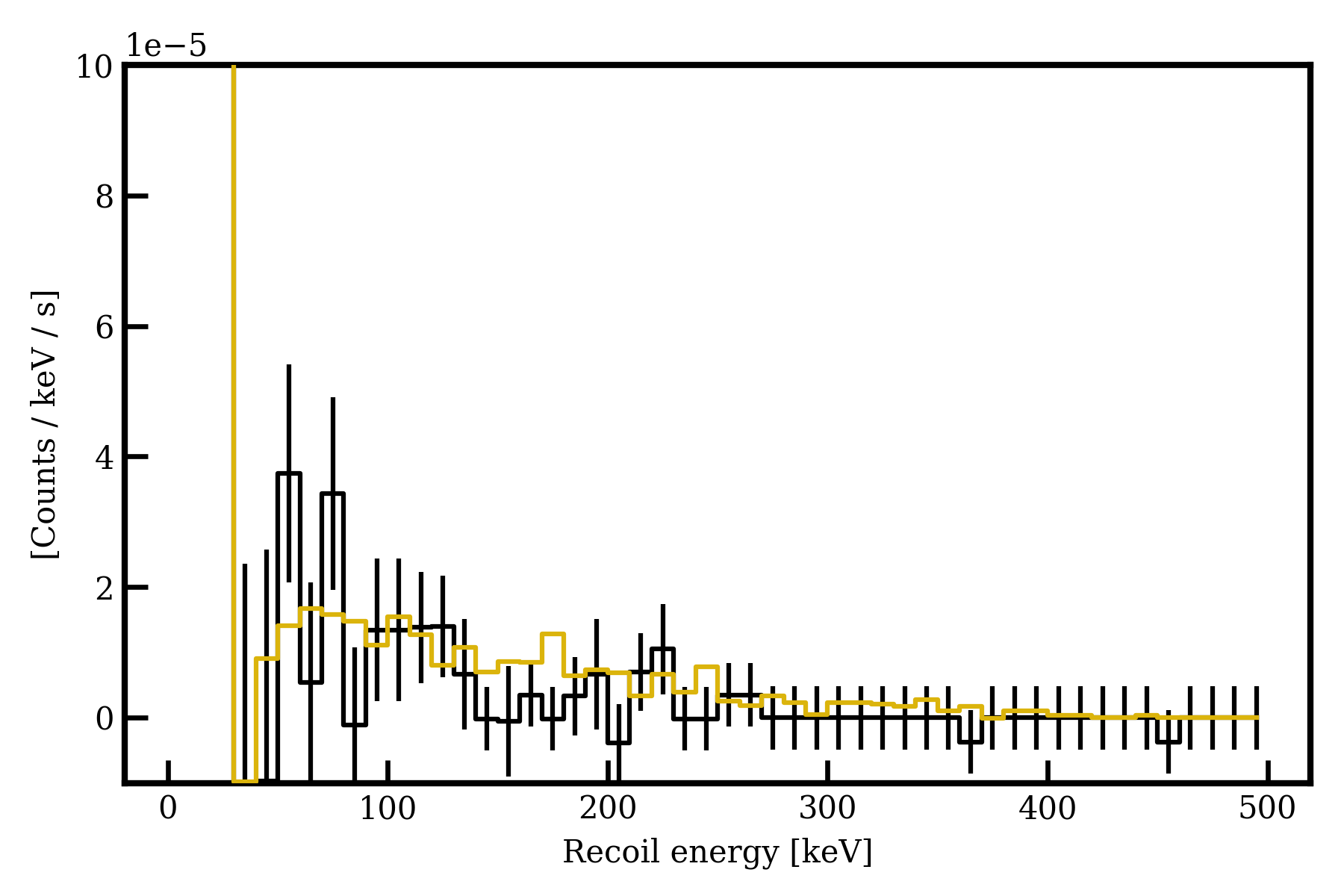}
\caption{\label{fig:proton_recoil_subtracted_high_energy} The gamma-subtracted recoil energy spectrum in the HGPC (black); showing the high energy contribution from the 390~keV neutron population. The expected spectrum from Monte Carlo simulation is shown in gold.}
\end{figure}

Along with the main 24~keV neutron population, the higher energy population at 379~keV is also evident in the HGPC data. Fig.~\ref{fig:proton_recoil_subtracted_high_energy} shows the recoil spectrum for neutrons, obtained identically as described above, out to 500~keV, compared to the \textsc{Geant4} simulation (with no free parameters). The total rate of events with 50-500~keV recoil energy in data is 1.66~mHz, whereas a rate of 2.21~mHz is expected from simulation. In this case, the discrepancy can be attributed to uncertainty in the neutron production cross section in the photonuclear process at this gamma energy. While the 0.41~mb cross section from ENDF/B-VIII.0 was used for simulation, lower values have been reported at 0.26~mb by Robinson~\cite{AR2016c} and 0.25~mb by Arnold \textit{et al.}~\cite{CA2012a}; our data is consistent with a cross section between these and the ENDF/B-VIII.0 value. This is encouraging for monoenergetic neutron calibration applications with the main 24~keV population from this source, as the data suggests that the higher energy background is even smaller than initially predicted. While simulations predict the higher energy neutron population to contribute 5\% of the neutron flux exiting the source opening, 0.75 times as much flux as predicted was observed for this population; combined with 2.16 times higher than expected flux of the main low energy population this amounts to only a 2\% contamination of the neutron flux by the 379~keV photoneutrons.

\begin{figure}[ht]
\includegraphics[width=\linewidth]{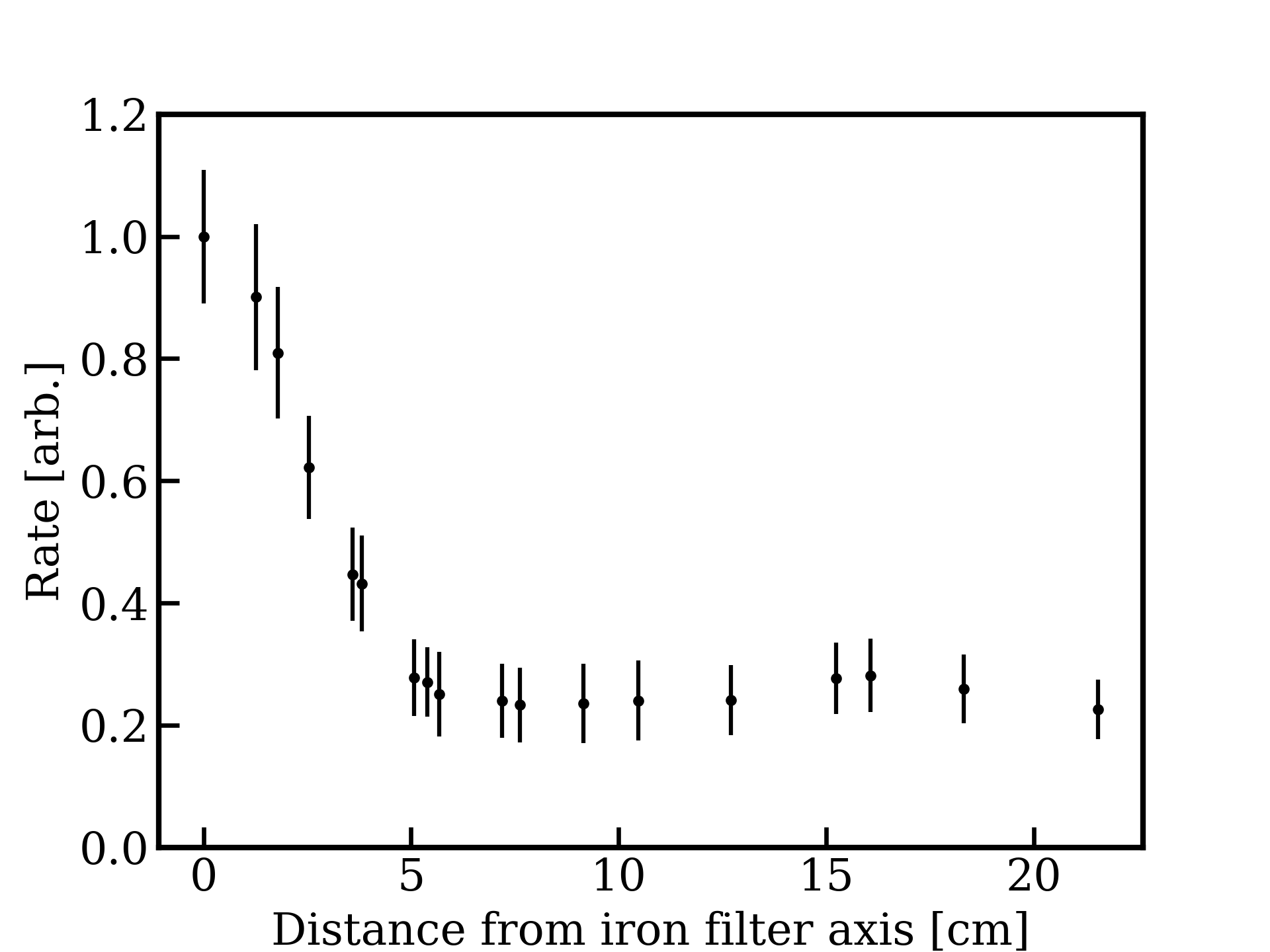}
\caption{\label{fig:proton_recoil_collimation} The pulse rate in the HGPC versus distance to the neutron beam axis, normalized to the on-axis rate. Eror bars represent statistical error multiplied by 10 for visibility.}
\end{figure}

In addition to characterizing the on-axis neutron spectrum, off-axis data was taken with the HGPC to measure the neutron beam profile. Since data taking with the HGPC is relatively slow, and preservation of 100\% of the neutron flux is not required to study the collimation, a slightly different strategy is employed to reduce gamma background. Data was only taken without the borated PE plug in place, and a stricter rise time cut accepting pulses between 0.6~$\mu$s and 1~$\mu$s is performed. The total rate of remaining pulses corresponding to proton recoil energies less than 50~keV is calculated on-axis and at varying off-axis positions, all with the center of the HGPC 6~cm higher than the face of the iron filter. Figure~\ref{fig:proton_recoil_collimation} shows how the rate drops with distance from the iron filter axis. At greater than 5~cm from the iron filter axis, no portion of the HGPC is directly above the iron rod, and the rate drops to a baseline of about 1/4 the on-axis rate. It is expected that this flat baseline is dominated by gammas. The lack of off-axis neutrons is valuable for reducing backgrounds from accidental coincidences and degraded neutrons in a calibration setup.

\begin{figure}[ht]
\includegraphics[width=\linewidth]{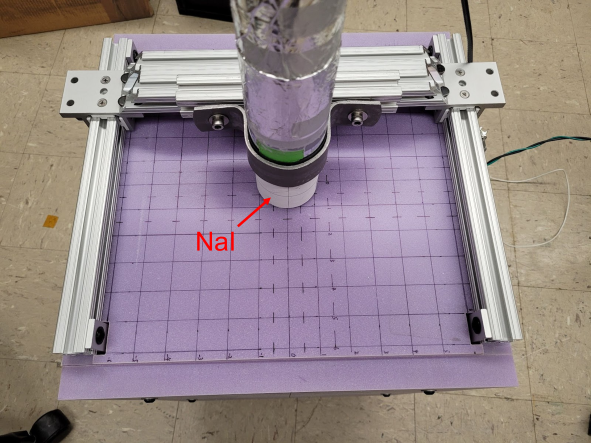}
\caption{\label{fig:NaI_setup} Photo of the NaI detector in position for measurement of the gamma spectrum escaping the source assembly.}
\end{figure}

\begin{figure}[ht]
\includegraphics[width=\linewidth]{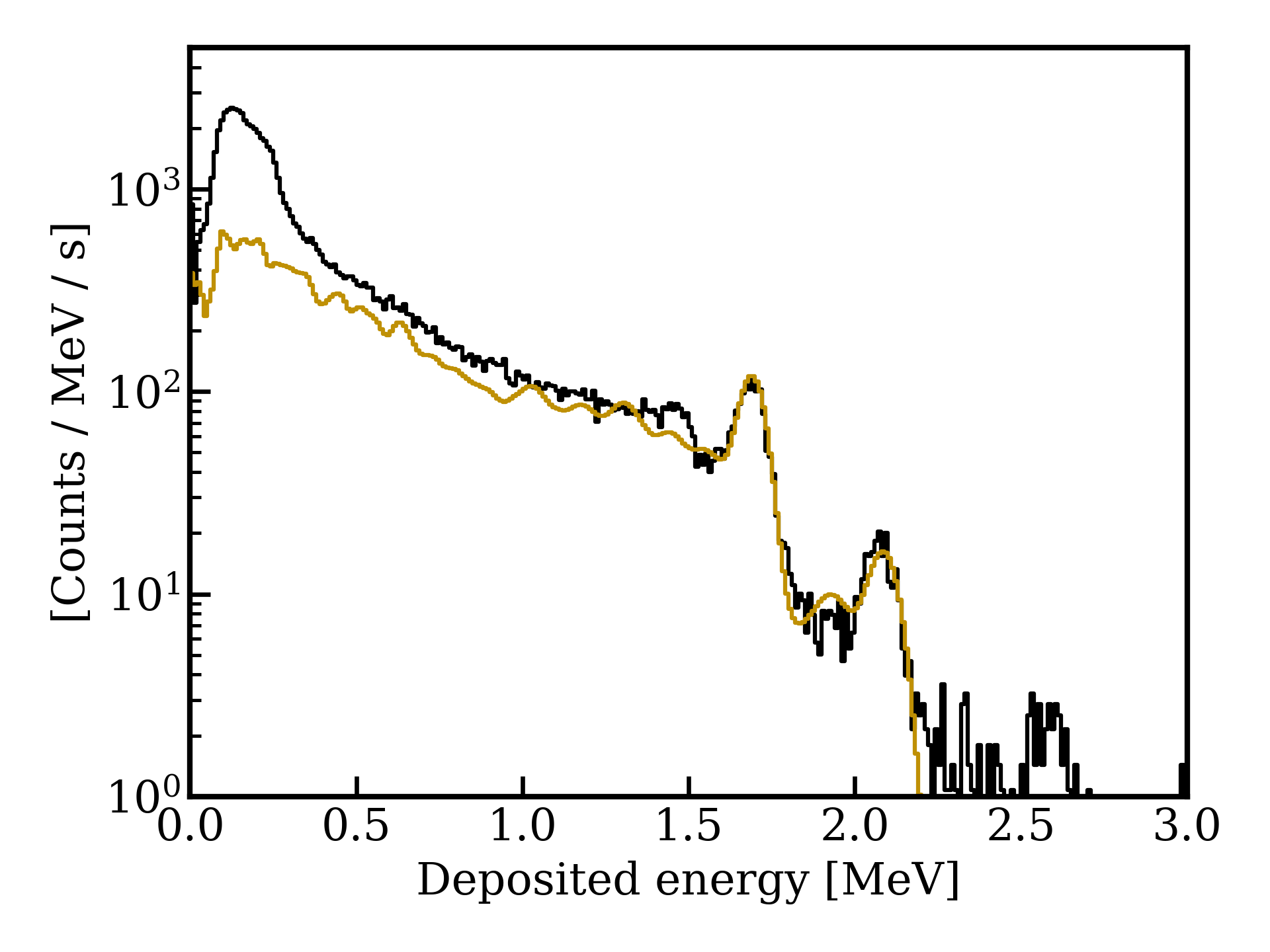}
\caption{\label{fig:NaI_spectrum} Energy spectrum in the NaI detector, data (black) and simulation (gold).}
\end{figure}

A NaI detector (Bicron Model 2M2/2) with 5.08~cm diameter and 5.08~cm length was used to measure the gamma background from this photoneutron source, as shown in Fig.~\ref{fig:NaI_setup}. Energy reconstruction with the NaI detector was calibrated using $^{241}$Am, $^{57}$Co, $^{133}$Ba, $^{137}$Cs, $^{22}$Na and $^{228}$Th sources. The NaI detector was placed above the iron filter, and its distance to the source was varied. As shown in Fig.~\ref{fig:NaI_spectrum}, the peaks due to full absorption of the 1691~keV and 2091~keV $^{124}$Sb gamma energies are observed with similar strength in data as in \textsc{Geant4} simulation at sufficient distance from the source surface, in this case with center of the NaI detector 35.1~cm above the top borated PE surface of the source assembly. At 52.7~cm, the furthest distance measured, the rate calculated from a gaussian fit to the 1691~keV peak is 94$\pm$3\% of the simulated rate, while the observed rate of the 2091~keV peak is 91$\pm$6\% of the simulated rate. At closer distances, the observed 1691~keV peak drops relative to the simulation, to 79$\pm$2\% of the simulated rate at the nearest distance of 2.3~cm. This discrepancy in the near-field gamma rate may indicate some shortcomings in the simulated geometry. On the other hand, the 2091~keV peak is consistent between data and simulation at the closest distances. Taking the ratio of observed to expected rate in the 1691~keV photoabsorption peak at the closest NaI distance to scale the full simulated gamma flux yields 213$\pm$6 gammas per cm$^2$ per second emitted from the exit of the  source assembly. Along with the HGPC neutron measurements, this indicates 2.5 neutrons in the desired 20-25~keV range are emitted per 100 gammas from the source opening. This is 2.73 times the simulated ratio, and represents a roughly 250-fold improvement over the bare SbBe source. 

%% file: Section_text/neutron_tagging.tex
A typical nuclear recoil energy calibration setup consists of a monoenergetic neutron source, a target detector to be calibrated, as well as a backing detector that can tag the angle of the scattered neutron to determine the fraction of energy deposited in the target. Fast detector response is desirable, as time-of-flight measurement can reduce backgrounds from gammas and accidental coincidences. A $^3$He proportional counter nested in polyethylene has a slow response due to the thermalization and capture process, further broadening the $\sim\mu$s resolution of the ionization signal. The HGPC also is relatively slow, and its low detection efficiency is prohibitive given the small solid angles required to perform a precise recoil angle calibration. Therefore, we investigate the potential of tagging neutrons with an Eljen-301 liquid scintillator detector.

The Eljen-301 liquid scintillator (LS) detector consists of a cylindrical cell with 12.7~cm diameter and 12.7~cm height filled with Eljen-301 LS, viewed by a Hamamatsu R877-100 photomultiplier tube (PMT). Light yield for low energy nuclear recoils in Eljen-301 has been studied by C. Awe et al (\cite{Awe:18prc}), including a measurement of the light yield relative to that of electronic recoils for recoil energies of 17.2~keV$_{nr}$ at 12.3\%. Given the 2.1 photoelectrons/keV$_{ee}$ observed in that setup, a signal of about 6 photoelectrons is expected for 24~keV neutrons transferring most of their energy into proton recoils in the LS. Our module consists of a significantly larger cell, and the PMT has a super-bialkali cathode rather than the ultra-bialkali cathode used in Ref.~\cite{Awe:18prc}, with a slightly smaller quantum efficiency (35\% with super-bialkali vs. 40\% with ultra-bialkali). Given the small expected signal size, it is a priori unclear that the LS detector will demonstrate significant sensitivity to neutrons below 24~keV.

\begin{figure}[ht]
\includegraphics[width=\linewidth]{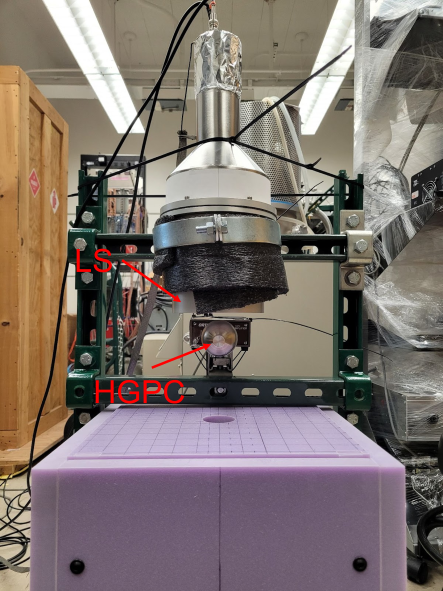}
\caption{\label{fig:SbBe_LS_setup} The setup for measuring SbBe neutron tagging with EJ-301 LS.}
\end{figure}

To test our Eljen-301 LS detector module's ability to detect low energy neutrons, the HGPC was employed to tag neutrons from the source incident on the LS detector, and a coincident signal in the LS detector was searched for. The HGPC was placed directly above the iron rod of the source assembly, and the LS detector was placed above the HGPC, as shown in Fig.~\ref{fig:SbBe_LS_setup}. Pulses in the HGPC triggered data acquisition of both the HGPC and the LS detector. Pulses in the LS detector in coincidence with neutron events identified in the HGPC could indicate a successfully tagged neutron. Coincident pulses could also arise from gammas; data with the removable borated PE plug was taken to quantify this background.

\begin{figure}[ht]
\includegraphics[width=\linewidth]{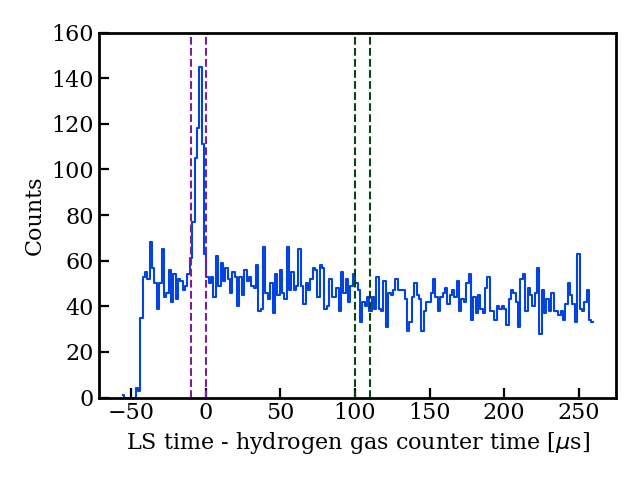}
\caption{\label{fig:delay_times} Distribution of time difference between LS pulses and HGPC pulses (blue), with no borated PE plug over the source. Purple dashed lines indicate the coincidence window, while green dashed lines indicate a non-coincidence window used to quantify accidental coincidence rate.}
\end{figure}

\begin{figure}[ht]
\includegraphics[width=\linewidth]{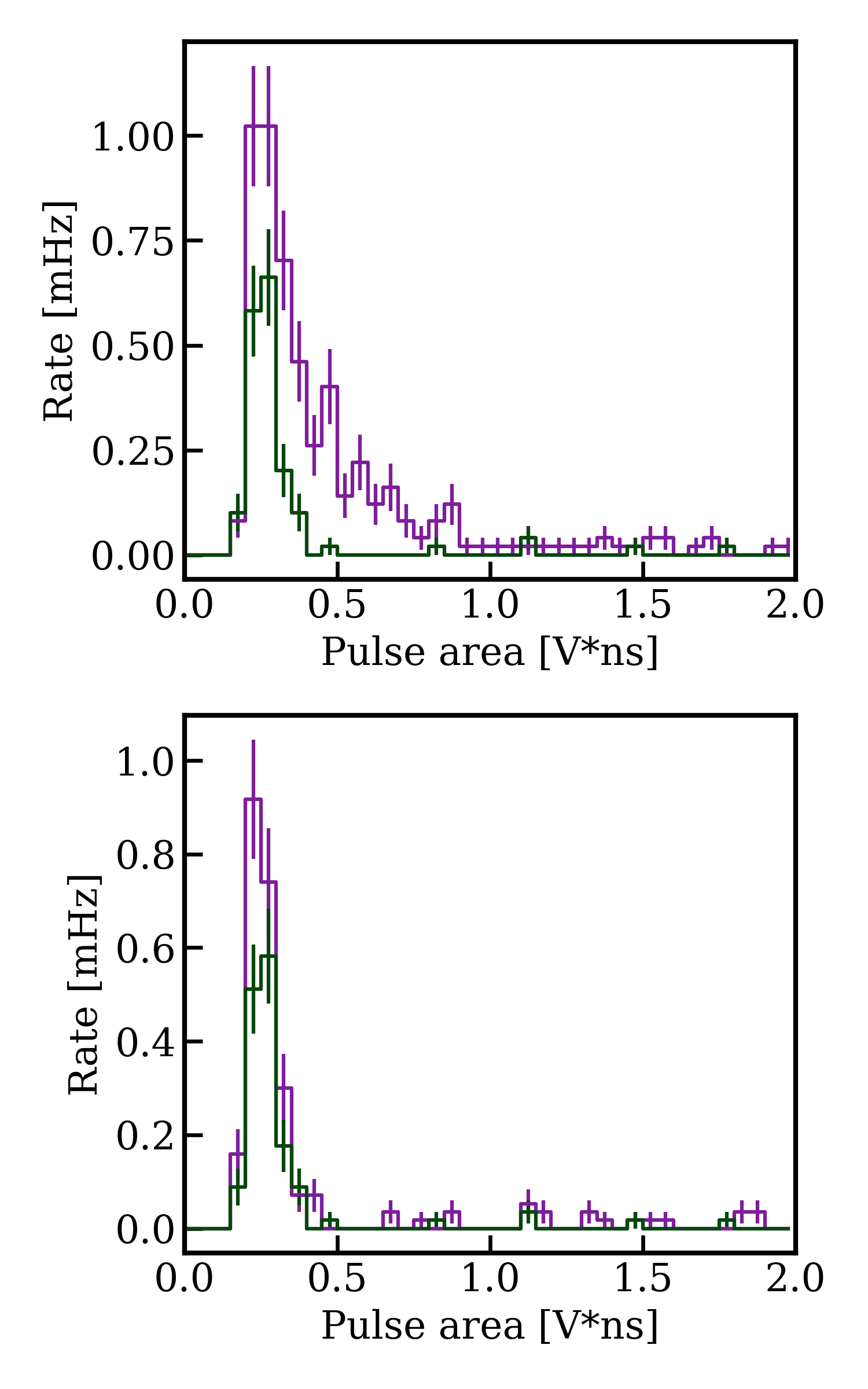}
\caption{\label{fig:LS_spectra} Top: pulse area spectra in the LS detector in (purple) and out (green) of coincidence with the HGPC, with no borated PE plug. Bottom: same, with borated PE plug.}
\end{figure}

As with the analysis of the neutron beam profile, a tighter rise time cut accepting pulses with rise time between 0.6~$\mu$s and 1~$\mu$s was applied to identify neutron events in the HGPC, since this analysis does not rely on accepting 100\% of the entire neutron flux. The distribution of time difference between pulses in the LS detector and the HGPC shows a clear peak from -10~$\mu$s to 0~$\mu$s, which is taken to define the coincidence window (see Fig.~\ref{fig:delay_times}). The spectrum of the coincident pulses in the LS detector is shown in Fig.~\ref{fig:LS_spectra}, along with the spectrum in a non-coincident window of (200,~210)~$\mu$s to compare the contribution of accidental coincidences. A clear excess of coincident pulses over non-coincident pulses at small area is observed, and is greatly diminished in data with the borated PE plug added, demonstrating that the LS detector is sensitive to the SbBe photoneutrons.

\begin{figure}[ht]
\includegraphics[width=\linewidth]{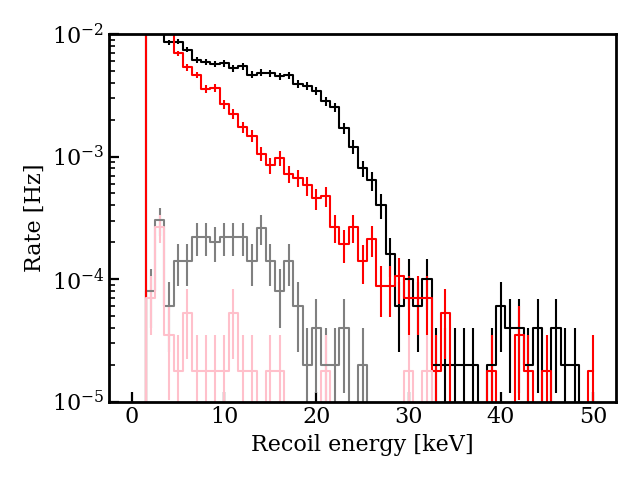}
\caption{\label{fig:proton_recoil_tagged_spectra} The observed energy spectrum in the HGPC, for all events without borated PE plug (black), events without borated PE plug with a tagged LS coincidence (gray), all events with borated PE plug (red), and events with borated PE plug with a tagged LS coincidence (pink).}
\end{figure}

\begin{figure}[ht]
\includegraphics[width=\linewidth]{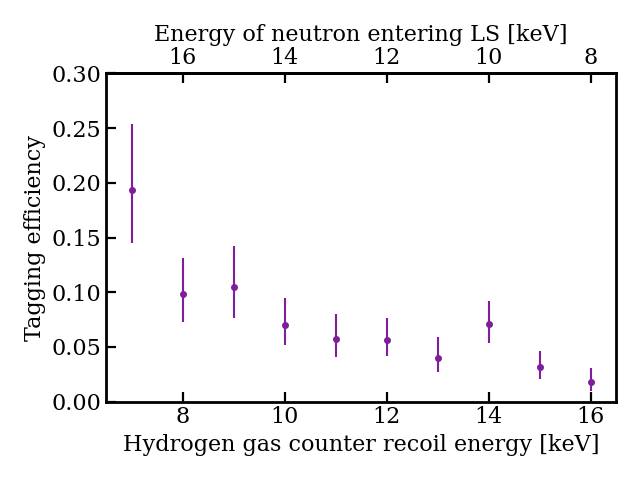}
\caption{\label{fig:tagging_efficiency} The efficiency of the LS module to tag neutrons as a function of recoil energy in the HGPC. The upper axis shows the incident neutron energy in the LS, assuming initial neutron energy of 24~keV in the HGPC.}
\end{figure}

Detection efficiency of the SbBe photoneutrons with the LS can be calculated by the rate of HGPC neutron events that have a coincident pulse in the LS. The full HGPC spectra after the rise time cut described above, with and without the borated PE plug, along with the spectra of events tagged by the LS are shown in Fig.~\ref{fig:proton_recoil_tagged_spectra}. Here, an LS tag is defined by the presence of an LS pulse in the coincident window with pulse area between 0.3 and 1.5 V*ns. As with the HGPC analysis of the neutron spectrum, the borated PE plug data is used to subtract gamma background from the full spectrum. Dividing the background-subtracted tagged spectrum from the background-subtracted full spectrum yields the tagging efficiency as a function of recoil energy in the HGPC shown in Fig.~\ref{fig:tagging_efficiency}. Here, the scaling of HGPC recoil energies that provided the best fit to the SbBe photoneutron spectrum discussed in Sec.\ref{sec:Source-Characterization} is assumed. The upper axis displays the corresponding energy of the neutron incident on the LS, which assumes the recoil energy is subtracted from monoenergetic 24~keV neutrons. Recoil energies below 7~keV are not reliably probed by the HGPC, since the spectrum at such small pulses becomes dominated by noise, and recoil energies higher than 16~keV are at angles too high to enter the LS detector in this configuration. As expected, the tagging efficiency increases with neutron energy, with a peak of 16\% for events with 7 keV recoil energy in the HGPC (17 keV incident on the LS detector). The PMT sample in this LS detector module offered poor gain, and did not exhibit a clear single photoelectron peak. It is expected that improvement in performance of the PMT coupled with the Eljen-301 cell could further boost the tagging efficiency of the LS module.

The EJ-301 module does demonstrate sensitivity to SbBe photoneutrons below 24~keV, and therefore can be paired with the source described herein to perform detector calibrations at small recoil energies. However, the detection efficiency is somewhat small, and for precise recoil angle tagging the solid angle of the module will also necessarily be small. The recoil energy resolution is directly related to the subtended polar angle of the backing detector as viewed from the target detector, and therefore quadratically related to solid angle. At small recoil angles especially, achieving reasonable energy resolution could imply a solid angle that results in an unacceptably low data rate. An improved geometry is a ring, placed around the axis of the source beam at the desired polar angle relative to the target. This increases the solid angle, and the scaling with energy resolution becomes linear. Such a geometry could be realized by an arrangement of multiple LS detectors. In addition, a backing detector based on neutron capture on $^6$Li with a ring shape has been designed in Ref.~\cite{Biekert:22nima}. Though the detector is much slower, implying more accidental coincidence background, it is mostly gamma-blind and is an especially valuable alternative to Eljen-301 for calibration at small recoil angles/energies where data accumulation could be unacceptably slow with the LS detector.

%% file: Section_text/conclusion.tex
The SbBe photoneutron source assembly described here provides a portable tool for calibration of detector response to low energy nuclear recoils. It achieves a high neutron flux by using a high activity $^{124}$Sb source activated at a nuclear reactor, and a high ratio of neutrons to gammas emitted due to the overlap of the SbBe photoneutron spectrum with a dip in the neutron cross section in iron. Measurements with a NaI detector indicate a flux of 213$\pm$6 gammas per cm$^2$ per second exiting the iron filter face. HGPC characterization of the source demonstrates a flux of the main neutron population approximately twice the predicted value, corresponding to 5.36$\pm$0.20 neutrons per cm$^2$ per second between 20 and 25 ~keV exiting the hole in the borated PE that collimates the neutron beam, while the secondary population at 379~keV is consistent with simulations considering the uncertainty in the photonuclear cross section, and only contributes 2\% to the total neutron flux from the iron filter. The gamma background in a low energy nuclear recoil calibration can also be quantified by temporarily plugging the iron filter with borated polyethylene.

Along with the novel low energy neutron source assembly, detection of the neutrons is demonstrated using Eljen-301 liquid scintillator. This provides a backing detector option for tagging recoil angles that is complementary to a detector based on neutron capture like the one described in Ref.~\cite{Biekert:22nima}. The latter delivers superior detection efficiency compared to what is measured in the LS module here, and it is insensitive to gamma backgrounds. However, the faster LS detector can better reduce various backgrounds with a tighter coincidence window. Where especially precise recoil angle tagging is desired, a ring-shaped backing detector may be required for acceptable geometric efficiency. This can be achieved with the detector based on neutron capture mentioned above, or by arranging multiple LS detectors in a ring shape. The pairing of this photoneutron source with backing detectors presents a path for sub-keV nuclear recoil calibration of detectors such as those searching for low mass dark matter or CEvNS.